\documentclass[showpacs,amsmath,amssymb]{revtex4}

\usepackage{graphicx}
\usepackage{dcolumn}
\usepackage{bm}
\begin{document}

\title{Josephson effect in superconducting constrictions with hybrid SF electrodes:
peculiar properties determined by the misorientation of magnetizations}

\author{A. V. Zaitsev}
\affiliation{Institute of Radioengineering and Electronics of
the Russian Academy of Sciences, 125009 Moscow, Russia} 

\begin{abstract}
Josephson current in SFcFS junctions with arbitrary transparency of the
constriction (c) is investigated. The emphasis is done on the analysis of
the supercurrent dependencies on the misorientation angle $\theta $ between
the in-plane magnetizations of diffusive ferromagnetic layers (F). It is
found that the current-phase relation $I(\varphi )$ may be radically
modified with the $\theta $ variation: the harmonic $I_{1}\sin \varphi $
vanishes for definite value of $\theta $ provided for identical orientation
of the magnetizations ($\theta =0$) the junction is in the $"\pi"$ state.
The Josephson current may exhibit a nonmonotonic dependence on the
misorientation angle both for realization of $"0 "$ and $"\pi "$ state at $\theta =0$. 
We also analyze the effect of exchange field induced enhancement
of the critical current which may occur in definite range of $\theta $.
\end{abstract}

\maketitle

The dc Josephson effect in junctions with ferromagnetic interlayers\ (F)
exhibit exchange field induced\ remarkable features,\ namely, the transition
from a $"0"$ state to a $"\pi "$ state of junctions \cite{Bul,Buz82},\
enhancement of the critical Josephson current\ \cite{V1,V2}\ and others.
These features have been intensively studied in recent years theoretically
and experimentally (see Reviews \cite{GKI04,Ryaz,Buz05} and Refs. therein).
In spite of a lot of works devoted to this problem there are several
important ones which require additional investigation. The purpose of the
present work is the study of one of these problems. Namely we investigate
the Josephson current and peculiar properties of\ SFcFS junctions (c denotes
a constriction)\ with arbitrary misorientation angle $\theta $ between the
in-plane oriented magnetizations of the ferromagnetic metal (F) layers. This
investigation is important because the supercurrent\ is very sensitive to
the mutual orientation of magnetizations and, besides, because the variation
of $\theta $ gives additional\ opportunity to switch experimentally between
the $"0"$ \ and $"\pi "$ states. In the considered junctions the
constriction is an insulator characterized by the arbitrary transparency\ or
short (in comparison with the coherence length) diffusive metal channel.\
For the F layers the dirty limit, i.e. diffusive regime of electron
transport is assumed.\ Note that for the case of ballistic transport through
the F layers of quantum S-FIF-S contacts with arbitrary misorientation angle 
$\theta $ between the magnetizations in the F layers the Josephson current
has been studied by Barash \textit{et al.} \cite{BBK02}. In the present
paper we investigate the angle $\theta $-dependence of different peculiar
properties of the supercurrent in SFcFS junctions. We show that variation of 
$\theta $ may result in the\ transition between the $"0"$ and $"\pi "$
states and the disappearance of $I_{1}\sin \varphi $ harmonic of the
supercurrent for certain value of $\theta $. Effect of exchange field
induced enhancement of the critical current and its nonmonotonic dependence
on the angle $\theta $ are also studied.

As a model of the constriction, we consider an aperture of a small radius in
a thin impenetrable screen dividing two different electrodes.\ The
constriction is supposed to include a barrier with the transparency $%
D=D(\vartheta ),$ where $\vartheta $ is the angle between the electronic
trajectory and normal\ to the junction plane. Consider first the case of
ballistic constriction of small size\ assuming that its radius is small in
comparison with $l_{F},v_{F}/\Delta ,\;d_{F},$ where $l_{F}$ \ and $v_{F},\;$%
are\ the mean free path and Fermi velocity in the F layers, $d_{F}\;$is
their thickness, $\Delta \;$is the energy gap in the superconductors. Under
these assumptions the current may be expressed through the Matsubara Green's
functions of the F layers\ $\mathrm{\hat{G}}_{1,2}\;$as follows\ \cite
{Z84,ZA,Naz}: 
\begin{gather}
I(\varphi )=\frac{8\pi T}{eR_{0}}\sum_{n=0}^{\infty }\left\langle J(\omega
_{n})\right\rangle ,  \label{I1} \\
J=\frac{D}{4}\mathrm{Tr}\hat{\sigma}_{0}\otimes \hat{\tau}_{3}\frac{\mathrm{%
\hat{G}}_{-}\mathrm{\hat{G}}_{+}}{(1-D)\hat{1}+D\mathrm{\hat{G}}_{+}^{2}}=%
\frac{D}{4}\mathrm{Tr}\hat{\sigma}_{0}\otimes \hat{\tau}_{3}\frac{[\mathrm{%
\hat{G}}_{2},\mathrm{\hat{G}}_{1}]_{-}}{2(2-D)\hat{1}+D[\mathrm{\hat{G}}_{2},%
\mathrm{\hat{G}}_{1}]_{+}},  \notag
\end{gather}
where $\omega _{n}=\pi T(2n+1)\;$is\ the Matsubara frequency, $\mathrm{%
\hat{G}}_{\pm }=(\mathrm{\hat{G}}_{2}\pm \mathrm{\hat{G}}_{1})/2,\;[\mathrm{%
\hat{G}}_{2},\mathrm{\hat{G}}_{1}]_{\pm }=\mathrm{\hat{G}}_{2}\mathrm{\hat{G}%
}_{1}\pm \mathrm{\hat{G}}_{1}\mathrm{\hat{G}}_{2},\;R_{0}=4\pi ^{2}\hbar
^{3}/(e^{2}p_{F}^{2}\mathit{A})\;$is the Sharvin resistance of the contact,$%
\;\mathit{A\;}$being the area of the contact,\ angular brackets denote
angular-averaging$\;\left\langle (...)\right\rangle =2\int_{0}^{1}(...)\cos
\vartheta d\cos \vartheta ,\;\hat{\sigma}_{k}\;$and $\hat{\tau}_{k}$\ are
the Pauli ($k=1,2,3$)\ or unite ($k=0)\;$matrices in the spin and
particle-hole spaces, respectively,\ $\hat{1}=\hat{\tau}_{0}\otimes \hat{%
\sigma}_{0}$ \ As was noted above we consider the case of dirty thin F
layers,\ i.e. their thickness is supposed to satisfy the conditions\ $%
l_{F}\ll d_{F}\ll (\hbar \mathit{D}_{F}/T_{c})^{1/2}$ where$\;\mathit{D}%
_{F}\;$is the diffusive coefficient in the F layers,\ $T_{c}\;$is the
critical temperature of the superconductors.\ For such layers the space
variation of the Green's functions is negligible therefore they\ are given
by the expression $\mathrm{\hat{G}}_{j}=\exp (i\varphi _{j}\hat{\rho}_{3}/2)%
\mathrm{\hat{g}}_{j}\exp (-i\varphi _{j}\hat{\rho}_{3}/2)$, where $\hat{\rho}%
_{3}=\hat{\tau}_{3}\otimes \hat{\sigma}_{0},\;\varphi _{j}\;$is\ the\ phase\
of the order parameter of the $j$th\ S-electrode and $\mathrm{\hat{g}}_{j}$
is determined by the following equation (for details of the derivation see,
e.g., \cite{VZK}): 
\begin{equation}
\lbrack \mathrm{\hat{g}}_{j},\omega \hat{\rho}_{3}+i\mathbf{h}_{j}\mathbf{%
\hat{S}}+\varepsilon _{b}\mathrm{\hat{g}}_{S}]_{+}=\hat{0}  \label{Eqg}
\end{equation}
where $\varepsilon _{b}=\hbar \left\langle D_{b}\right\rangle v_{F}/4d_{F}=%
\mathit{D}_{F}/2R_{b}\sigma _{F}d_{F},\;D_{b}\;$and $R_{b}$ are,
respectively, the transparency and the resistance per unite area of the S-F
interface,$\;\sigma _{F}\;$is the conductivity of the F metal, $\hat{g}_{S}\;
$is the Green's function of the superconductors at the S-F interfaces,$\;%
\mathbf{h}_{1,2\;}\;$are the exchange fields in the F layers, the matrix
vector$\;\mathbf{\hat{S}\;}$is defined as\ \cite{Maki,Alex}$\;\mathbf{\hat{S}%
=\hat{\sigma}}\otimes \hat{\tau}_{+}\mathbf{-}\hat{\sigma}_{2}\mathbf{\hat{%
\sigma}}\hat{\sigma}_{2}\otimes \hat{\tau}_{-},\;$where\ $\mathbf{\hat{\sigma%
}=(}\hat{\sigma}_{1},\hat{\sigma}_{2},\hat{\sigma}_{3})\mathbf{,\;}\hat{\tau}%
_{\pm }=(\hat{\tau}_{0}\pm \hat{\tau}_{3})/2.$ Note that the value of the
exchange fields,\ $\left| \mathbf{h}_{1,2}\right| =h$ is supposed to be
small with respect to $\hbar /\tau _{F}\;$and the Fermi energy of the F
layers, where$\;\tau _{F}=l_{F}/v_{F};\;$for considered here dirty layers $%
\hbar /\tau _{F}\gg \Delta $.\ We assume for simplicity that the S-F
interfaces are identical and that\ their transparency $D_{b}$ is small,
therefore the effect of the F layers on the superconductors is negligible.\
The only difference between the F layers is related with the orientations of
the exchange fields, $\mathbf{h}_{1,2},\;$which\ being parallel to the\
layers,\ make an angle $\theta $ with each other. In this paper we confine
ourselves to the case of conventional s-wave superconductors.\ Therefore\ to
carry out further calculations it is convenient \ to represent the current
with the use of the transformed Green's functions\ $\hat{G}_{j}=\exp
(i\varphi _{j}\hat{\rho}_{3}/2)\hat{g}_{j}\exp (-i\varphi _{j}\hat{\rho}%
_{3}/2)\;$where 
\begin{equation}
\hat{g}_{j}=\hat{U}\mathrm{\hat{g}}_{j}\hat{U}^{-1},\;  \label{Transf}
\end{equation}
$\hat{U}=(\hat{\sigma}_{0}\otimes \hat{\tau}_{+}+\hat{\sigma}_{2}\otimes 
\hat{\tau}_{-}).\;$Taking into account that particle (hole) projection
matrices, $\hat{\tau}_{+(-)}$ obey the relation $\hat{\tau}_{\alpha }\hat{%
\tau}_{\beta }=\hat{\tau}_{\alpha }\delta _{\alpha \beta }\;$it is easy to
check that $\hat{U}$ is the unitary Hermitian\ matrix, $\hat{U}$ $=\hat{U}%
^{-1}=\hat{U}^{+}.\;$From (\ref{Transf}), (\ref{Eqg})\ one can find the
representation for\ $\hat{g}_{j}\;$with the help of the projection matrices
in the spin space,$\;\hat{P}_{\pm }(\mathbf{n}_{j})=\frac{1}{2}(\hat{\sigma}%
_{0}\pm \mathbf{n}_{j}\mathbf{\hat{\sigma}}),\;$where\ the unite vectors$\;%
\mathbf{n}_{j}=\mathbf{h}_{j}/\left| \mathbf{h}_{j}\right| .\;$Taking into
account the relation$\;\hat{P}_{\alpha }(\mathbf{n})\hat{P}_{\beta }(\mathbf{%
n})=\hat{P}_{\alpha }(\mathbf{n})\delta _{\alpha \beta },\;$we get 
\begin{equation}
\hat{g}_{j}=\sum_{\alpha =\pm }\hat{g}_{\alpha }\otimes \hat{P}_{\alpha }(%
\mathbf{n}_{j})=\frac{1}{2}(\hat{g}_{+}+\hat{g}_{-})\otimes \hat{\sigma}_{0}+%
\frac{1}{2}(\hat{g}_{+}-\hat{g}_{-})\otimes \mathbf{n}_{j}\mathbf{\hat{\sigma%
}},  \label{gtr}
\end{equation}
where $\hat{g}_{\alpha }=g_{\alpha }\hat{\tau}_{3}+f_{\alpha }\hat{\tau}%
_{2},\;g_{\alpha }^{2}+f_{\alpha }^{2}=1,\;$%
\begin{equation}
g_{\alpha }=\frac{\tilde{\omega}_{\alpha }}{\xi _{\alpha }},\;\;f_{\alpha }=%
\frac{\Delta _{F}}{\xi _{\alpha }},\;
\end{equation}
$\xi _{\alpha }=\sqrt{\tilde{\omega}_{\alpha }^{2}+\Delta _{F}^{2}},\;\tilde{%
\omega}_{\alpha }=\omega +i\alpha h+\varepsilon _{b}g_{S}(\omega ),\;\Delta
_{F}=\varepsilon _{b}f_{S}(\omega ).\;$The form of the Green's function
representation\ (\ref{gtr})\ does not depend on the specific expressions for 
$\hat{g}_{\alpha };\;$it is valid for spatially-homogeneous\ orientation of
the exchange fields in the F layers. Note that\ being written via the
transformed matrices $\hat{G}_{j},\;$the expression for the current
coincides with (1).\ The advantage of the representation (\ref{gtr})\ is
related with significant simplification of the matrix structure of the
Green's functions which are given by sum of two terms determined by the
direct product of matrices in the particle-hole and spin spaces.\ With the
help of (\ref{gtr})\ we get 
\begin{equation}
\lbrack \hat{G}_{2},\mathit{\ }\hat{G}_{1}]_{\pm }=\frac{1}{2}\sum_{\alpha
,\beta =\pm }[\hat{G}_{2\alpha },\hat{G}_{1\beta }]_{+}\otimes \lbrack \hat{P%
}_{\alpha }(\mathbf{n}_{2}),\hat{P}_{\beta }(\mathbf{n}_{1})]_{\pm }+[\hat{G}%
_{2\alpha },\hat{G}_{1\beta }]_{-}\otimes \lbrack \hat{P}_{\alpha }(\mathbf{n%
}_{2}),\hat{P}_{\beta }(\mathbf{n}_{1})]_{\mp },  \label{c&ac}
\end{equation}
where$\;\hat{G}_{j\alpha }=g_{\alpha }\hat{\tau}_{3}+\exp (i\varphi _{j}\hat{%
\tau}_{3})f_{\alpha }\hat{\tau}_{2}$. Taking into account that 
\begin{eqnarray}
\lbrack \hat{P}_{\alpha }(\mathbf{n}_{2}),\hat{P}_{\beta }(\mathbf{n}%
_{1})]_{+} &=&\frac{1}{2}[1+\alpha \beta \mathbf{n}_{2}\mathbf{n}%
_{1}+(\alpha \mathbf{n}_{2}\mathbf{+}\beta \mathbf{n}_{1})\mathbf{\hat{\sigma%
}}],\;  \notag \\
\lbrack \hat{P}_{\alpha }(\mathbf{n}_{2}),\hat{P}_{\beta }(\mathbf{n}%
_{1})]_{-} &=&\frac{i\alpha \beta }{2}[\mathbf{n}_{2}\times \mathbf{n}_{1}]%
\mathbf{\hat{\sigma},}  \label{c&ac1}
\end{eqnarray}
after some calculations we obtain the following expression for the current 
\begin{equation}
I(\varphi )=\frac{8\pi T\sin \varphi }{eR_{0}}\sum_{n=0}^{\infty
}\left\langle \frac{A(\varphi )}{B_{0}-[A_{0}+A(\varphi )]\sin ^{2}\left( 
\frac{\varphi }{2}\right) }\right\rangle \equiv \left\langle \mathrm{I}%
(\varphi ,D)\right\rangle ,  \label{IM}
\end{equation}
where\ $\varphi =\varphi _{2}-\varphi _{1},\;A(\varphi )=A_{0}+A_{1}\sin
^{2}\left( \frac{\varphi }{2}\right) ,\;$%
\begin{eqnarray*}
A_{0} &=&2D\left[ 2\cos ^{2}\left( \frac{\theta }{2}\right) \mathop{\rm Re}%
f^{2}+\sin ^{2}\left( \frac{\theta }{2}\right) \left| f\right| ^{2}(2+Dq)%
\right] , \\
A_{1} &=&-4D^{2}\left| f\right| ^{4},\;B_{0}=\left[ 2+D\sin ^{2}\left( \frac{%
\theta }{2}\right) q\right] ^{2},\;\;q=\left| g\right| ^{2}+\left| f\right|
^{2}-1,
\end{eqnarray*}
here$\;g\equiv g_{+}(\omega _{n}),\;f\equiv f_{+}(\omega _{n}).\;$Eq. (\ref
{IM})\ is significantly simplified for the cases of parallel (p) and
antiparallel (a) magnetizations: 
\begin{equation}
I=\frac{8\pi T\sin \varphi }{eR_{0}}\sum_{n=0}^{\infty }\left\{ 
\begin{array}{l}
\mathop{\rm Re}\left\langle Df^{2}/[1-Df^{2}\sin ^{2}\left( \frac{\varphi }{2%
}\right) ]\right\rangle ,\;p \\ 
\left\langle 2D\left| f\right| ^{2}/[2+D(\left| g\right| ^{2}+\left|
f\right| ^{2}\cos \varphi -1)]\right\rangle ,\;a
\end{array}
\right.   \label{Ipa}
\end{equation}
Note that for the case $\theta =0\;$Eq.(\ref{Ipa}) reduces to the result
obtained in Ref.\cite{GKF}. In general case the misorientation angle
dependence of the current is rather nontrivial if the transparency $D\;$is
not small: it does not reduce to the expression\ (if $\mathbf{h}%
_{1}\nparallel \mathbf{h}_{2}$) \cite{note} 
\begin{equation}
I(\varphi )=I^{(p)}(\varphi )\cos ^{2}\left( \frac{\theta }{2}\right)
+I^{(a)}(\varphi )\sin ^{2}\left( \frac{\theta }{2}\right) .  \label{TunJ}
\end{equation}
Simple form for the $\theta -$dependence (\ref{TunJ})\ is valid only in the
limit of small transparencies\ (when$\;I^{(p,a)}(\varphi )=I^{(p,a)}\sin
\varphi $); for this case it was first obtained in Ref.\cite{V1}.

For numerical analysis of the Josephson current we apply the obtained
results assuming that superconductors are described by the BCS theory
therefore$\;g_{S}(\omega )=\omega /(\omega ^{2}+\Delta
^{2})^{1/2},\;f_{S}(\omega )=\Delta /(\omega ^{2}+\Delta ^{2})^{1/2}$. The
results of numerical calculations of $I(\varphi )\;$-dependences for a
junction with the transparency $D=0.5\;$are presented in Fig.1 for a set of
values of $\theta $. It shows that the function $I(\varphi )$ may strongly
deviate from the $\sin \varphi -$dependence and is strongly modified with
increasing angle $\theta $, especially\ in the vicinity of \ misorientation\
angle$\;\theta =\theta _{0\pi }\;$(temperature $T_{0\pi })\;$which\
corresponds to the transition between the $"0"$ and $"\pi "\;$states of
junction. Note that the model with the angle-independent transparency
corresponds to the case of quantum single-mode constriction with the
resistance $R_{0}=\pi \hbar /e^{2}$. At definite $\theta =\theta ^{\ast }\;$%
(temperature$\;T^{\ast }$) which with the accuracy higher than 5 percents
coincides with the value $\theta _{0\pi }$\ ($T_{0\pi })$ the harmonic $%
I_{1}\sin \varphi \;$of the current ($I(\varphi )=I_{1}\sin \varphi
+I_{2}\sin (2\varphi )+...)\;$vanishes; the sign of$\;I_{1}\;$is changed
with the variation of $\theta \;$(or temperature) near $\theta ^{\ast }$\ $%
(T^{\ast }).$ Living the detailed analysis of supercurrent temperature
dependences for a separate presentation\ we show here the typical phase
diagram of the junction (Fig.2)\ for not too small exchange fields (in
comparison with $\varepsilon _{b})$.\ Another interesting property of the
supercurrent is nonmonotonic dependences of the critical current $I_{c}=\max
\left| I(\varphi )\right| \;$as a function of the angle $\theta $.\ Note
that a similar behavior of the critical current for quantum S-FIF-S
junctions with the ballistic transport through the F layers has been
analyzed in \cite{BBK02}. We find that in junctions with diffusive F layers
the nonmonotonic dependence $I_{c}(\theta )$ (Fig.3) may occur for
realization of both $"\pi " $ and $"0"$ state for identical orientation of
the magnetizations (unlike the case studied in \cite{BBK02}). The results of
numerical calculations of the $I_{c}(h)$ dependences for a low transparency
junction at various angle $\;\theta $ are presented\ in Fig.4. It shows that
the ratio $I_{c}(h)/I_{c}(0)$ may exceed unity, i.e., the effect of the
exchange field induced enhancement of the critical current \cite{V1,V2} may
occur in some range of $h\;$if $\theta \;$exceeds definite value ($\pi
/4<\theta \leq \pi \;$for small $D$).

For the case of SFcFS junctions with a short diffusive constriction the
following expression for the current may be obtained with the use of the
Green's function technique \cite{ZA} 
\begin{equation}
I(\varphi )=\int_{0}^{1}\rho (D)\mathrm{I}(\varphi ,D)dD,  \label{ID}
\end{equation}
where $\mathrm{I}(\varphi ,D)\;$is determined by (\ref{IM})\ with $R_{0}=\pi
\hbar /e^{2}\;$being\ the resistance\ of a single mode constriction, and the
distribution of transparencies is determined by the function $\rho
(D)=(R_{0}/2R_{N})/D(1-D)^{1/2}\;$($R_{N}\;$is the normal state resistance
of the constriction)\ which coincides with the density function found by
Dorokhov \cite{Dor}. For the cases of parallel and antiparallel
magnetizations\ Eqs.(\ref{IM}),\ (\ref{ID}) reduce to 
\begin{equation}
I(\varphi )=\frac{16\pi T\sin \varphi }{eR_{N}}\sum_{n=0}^{\infty }\left\{ 
\begin{array}{c}
\mathop{\rm Re}F^{2}W\left( 1-2f^{2}\sin ^{2}\left( \frac{\varphi }{2}%
\right) \right) ,\;p \\ 
\left| F\right| ^{2}W(\left| g\right| ^{2}+\left| f\right| ^{2}\cos \varphi
),\;a
\end{array}
\right.  \label{IDpa}
\end{equation}
where $W(x)=(1-x^{2})^{-1/2}\arctan [(1-x)/(1+x)]^{1/2}$. Note\ that in the
case $\;\theta =0\;$Eq.(\ref{IDpa})\ reduces to the result obtained in Ref. 
\cite{GKF}; for $h=0$ Eqs.(\ref{IDpa})\ coincide\ and in the limit $%
\varepsilon _{b}\gg \Delta \;$ reproduce\ the Kulik-Omelyanchuk\ formula 
\cite{KO} for a diffusive ScS constriction. Numerical calculations show
(Fig.5) that the above discussed peculiar properties manifests itself also
in junctions with diffusive constriction.

In conclusion, we have developed a microscopic theory of supercurrent in
SFcFS junctions with the arbitrary transparency of the constriction and
diffusive electron transport in the F layers for arbitrary angle $\theta \;$
between the in plane exchange fields in the F layers. We showed that the
current-phase dependence may be strongly modified with the variation of $%
\theta .\;$If the junction is in the $"\pi "\;$state\ for identical
orientation of magnetizations ($\theta =0)$, the variation of $\theta \;$may
result in the transition between the $"0"$ and $"\pi "\;$states and
disappearance of the $I_{1}\sin \varphi \;$harmonic of the supercurrent at
definite value of $\theta $. Nonmonotonic dependence of the critical current
as a function of $\theta $ may occur for both $"\pi "\;$ and $"0 "\;$ state
of junction corresponding to the case of identical orientation of the
magnetizations if the transparency of the junction is not small.

This work was supported by the RFBR (Grant No. 04-02-16818-a), INTAS (Grant
No. 2001-0809) and ISTC (Grant No. 2369).

\newpage

\begin{figure}
\includegraphics[clip=true,width=0.6\textwidth]{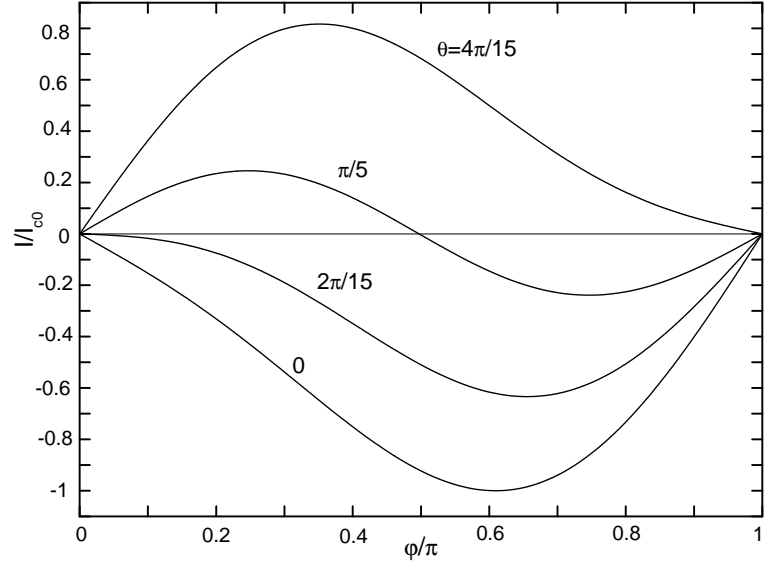}
\caption{Fig.1. Current-phase relations at various misorientation angle $\theta $ for
a SF-I-FS junction\ with the transparency of the barrier $D=0.5\;$at$%
\;h/\Delta =1.5,\;T/\Delta =0.15,\varepsilon _{b}=\Delta .\;$For $\theta
\leq \theta _{0\pi }=\pi /4\;(\theta _{0\pi }<\theta \leq \pi )\;$the
junction is in the $\ "\pi "\;$state ($"0"\;$state). For $\theta $ in the
vicinity of $\theta _{0\pi }\;$second harmonic $I_{2}\sin (2\varphi )\;$of
the current\ dominates,\ first harmonics $I_{1}\sin \varphi \;$vanishes at $%
\theta =\theta ^{\ast }\approx \theta _{0\pi }.$}
\label{fig1}
\end{figure}

\begin{figure}
\includegraphics[clip=true,width=0.6\textwidth]{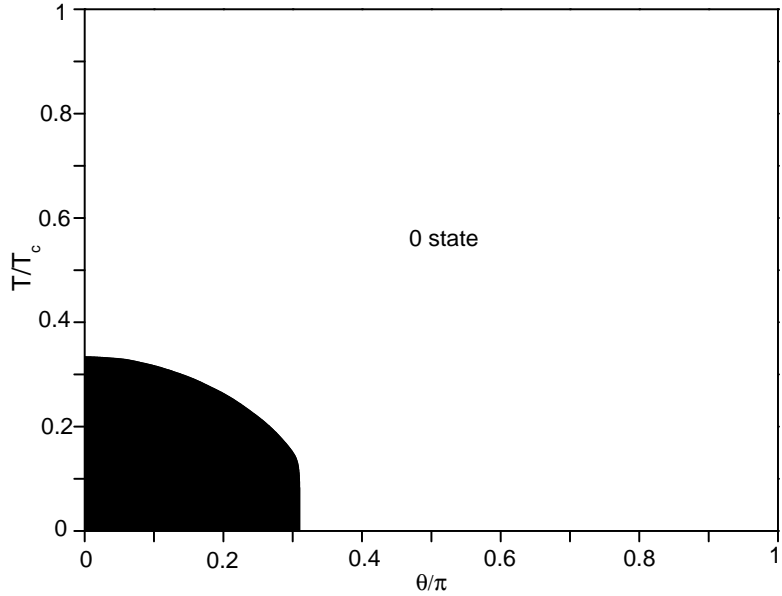}
\caption{Fig.2. The $(T,\theta )\;$phase diagram of a SF-I-FS junction with the
transparency of the barrier $D=0.5\;$and $h/\Delta =1.5,\;\varepsilon
_{b}=\Delta .\;$Black\ region\ corresponds to the $"\pi "\;$state of the
junction.}
\label{fig2}
\end{figure}

\begin{figure}
\includegraphics[clip=true,width=0.6\textwidth]{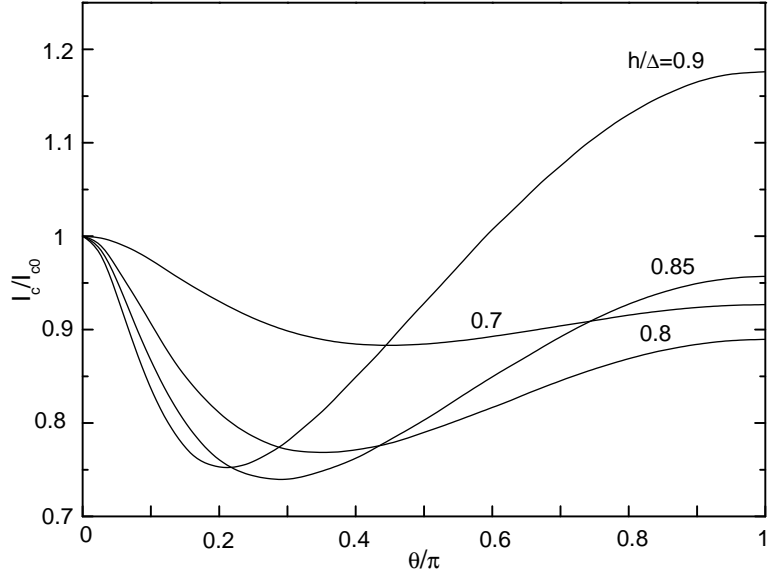}
\caption{Fig.3. Critical current as a function of the misorientation angle $\theta\;$
normalized to its value at $\theta =0\;$for a SF-I-FS junction with the
transparency of the barrier$\;D=0.5;$ $\varepsilon _{b}=\Delta ,\;T/\Delta
=0.01;\;$ these dependences correspond to $"0"$ state of the junction.}
\label{fig3}
\end{figure}

\begin{figure}
\includegraphics[clip=true,width=0.6\textwidth]{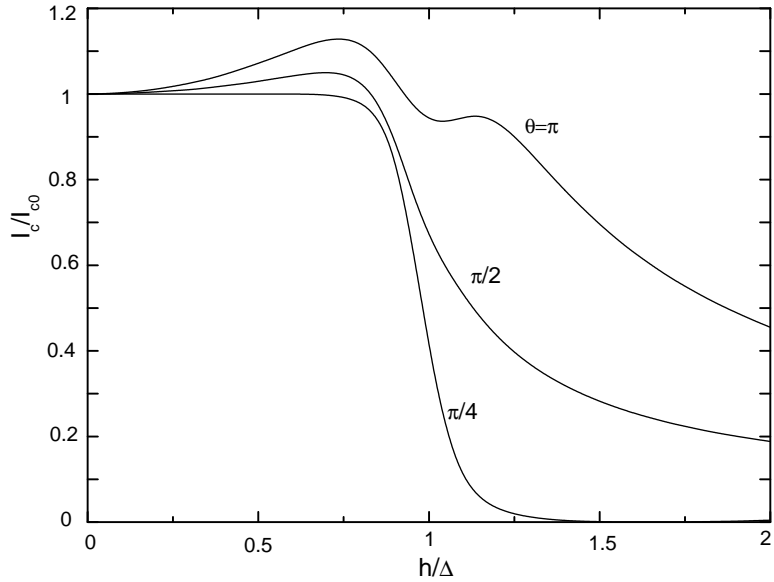}
\caption{Fig.4. Exchange field dependence of the critical current $I_{c}(h)\;($%
normalized to $I_{c}(0)\equiv I_{c0})$ at various misorientation angle $%
\theta \;$for SF-I-FS junction\ with the transparency of the barrier $%
D=0.1\; $at$\;T/\Delta =0.01,\varepsilon _{b}=\Delta .$}
\label{fig4}
\end{figure}

\begin{figure}
\includegraphics[clip=true,width=0.6\textwidth]{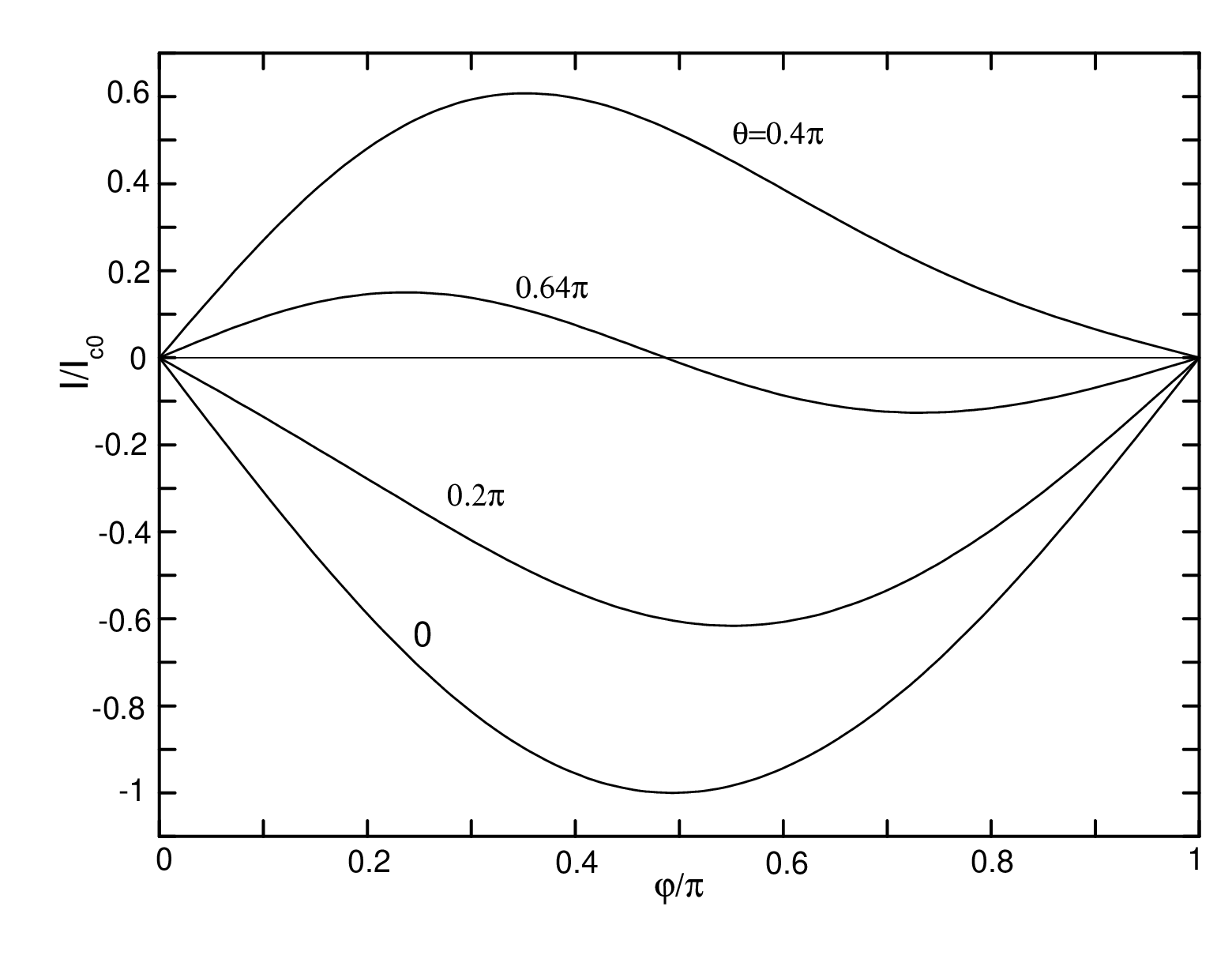}
\caption{Fig.5. Current-phase relations\ at various $\theta \;$for a diffusive
SF-c-FS junction$\;$at $h/\Delta =1.5,T/T_{c}=0.1,\varepsilon _{b}=\Delta
.\; $For $\theta \leq \theta _{0\pi }=0.64\pi \;(\theta _{0\pi }<\theta \leq
\pi )\;$the junction is in the $\ "\pi "\;$state\ ($"0"\;$state). For $%
\theta $ in the vicinity of $\theta _{0\pi }\;$second harmonic $I_{2}\sin
(2\varphi )\;$of the current\ dominates,\ first harmonics $I_{1}\sin \varphi
\;$vanishes at $\theta =\theta ^{\ast }\approx \theta _{0\pi }.$}
\label{fig5}
\end{figure}


\begin{thebibliography}{99}
\bibitem{note}  Note that in Ref.\cite{Ch02}\ the Josephson current in S$%
_{F} $cS$_{F}\;$junctions formed by two ''ferromagnetic superconductors'' (S$%
_{F}$) was studied on the basis of the Green's function approach. The
authors of this work arrived at the conclusion about the validity of formula
(\ref{TunJ}) for arbitrary transparency $D$ of the barrier in the
constriction. In reality this conclusion is not true if $\mathbf{h}%
_{1}\nparallel \mathbf{h}_{2}.$ The origin of the erronious conclusion is
related with that in the process of the derivation of the expression for the
current in Ref.\cite{Ch02} the authors used the identity (Eq.(5) of Ref.\cite
{Ch02}) which is not applicable for functions determined by the partitioned
matrices. It becomes clear from Eqs.(\ref{c&ac}), (\ref{c&ac1}) which show
that the expressions for $[\hat{G}_{2},\mathit{\ }\hat{G}_{1}]_{\pm }$, and,
as a consequence, the current depend not only on the the scalar product $(%
\mathbf{h}_{1}\mathbf{h}_{2})\;$ but also on the vector product $[\mathbf{h}%
_{1}\times \mathbf{h}_{2}]$.
\end{thebibliography}
\end{document}